\documentclass[floats,floatfix,showpacs,amssymb,prd,twocolumn,superscriptaddress,nofootinbib,nolongbibliography,reprint]{revtex4-1}

\usepackage{amssymb,amsmath,verbatim,mathtools,needspace,enumitem,etoolbox,graphicx,physics,microtype,afterpage,bigints,textcomp,gensymb,tabularx}

\usepackage[dvipsnames, usenames]{xcolor}
\definecolor{linkcolor}{rgb}{0.0,0.3,0.5}
\definecolor{dodgerblue}{HTML}{1E90FF}
\usepackage[unicode, colorlinks=true, linkcolor=linkcolor, citecolor=linkcolor, filecolor=linkcolor,urlcolor=linkcolor, pdfusetitle]{hyperref}
\usepackage[all]{hypcap}
\usepackage[T1]{fontenc}
\usepackage[utf8]{inputenc}
\usepackage{orcidlink}
\usepackage{tikz}

\newcommand{\ssim}{\mathchar"5218\relax\,}

\makeatletter
\newcommand*{\balancecolsandclearpage}{\close@column@grid \cleardoublepage \twocolumngrid}
\makeatother

\usepackage{lipsum}

\newcommand{\bham}{\affiliation{School of Physics and Astronomy \& Institute for Gravitational Wave Astronomy, University of Birmingham, \\ Birmingham, B15 2TT, United Kingdom}}
\newcommand{\milan}{\affiliation{Dipartimento di Fisica ``G. Occhialini'', Universit\'a degli Studi di Milano-Bicocca, Piazza della Scienza 3, 20126 Milano, Italy}}
\newcommand{\infn}{\affiliation{INFN, Sezione di Milano-Bicocca, Piazza della Scienza 3, 20126 Milano, Italy}}
\newcommand{\geneva}{\affiliation{ D\'epartement de Physique Th\'eorique \& Gravitational Wave Science Center (GWSC), Universit\'e de Gen\`eve, 24 quai Ernest Ansermet, 1211 Gen\`eve 4, Switzerland}}

\begin{document}

\title{
Inferring, not just detecting: metrics for high-redshift sources observed with third-generation gravitational-wave detectors
}

\author{Michele Mancarella\,\orcidlink{0000-0002-0675-508X}}
\email{michele.mancarella@unimib.it}

\milan \infn

\author{Francesco Iacovelli\,\orcidlink{0000-0002-4875-5862}}

\geneva \milan

\author{Davide Gerosa\,\orcidlink{0000-0002-0933-3579}}

\milan \infn \bham


\date{\today}

\begin{abstract}

The detection of black-hole binaries at high redshifts is a cornerstone of the science case of third-generation gravitational-wave interferometers. The star-formation rate peaks at $z\ssim 2$ and decreases by orders of magnitude by  $z\ssim 10$. Any confident detection of gravitational waves from such high redshifts would imply either the presence of stars formed from pristine material originating from cosmological nucleosynthesis (the so-called population III stars), or black holes that are the direct relics of quantum fluctuations in the early Universe (the so-called primordial black holes). Crucially, \emph{detecting} sources at cosmological distances does not imply \emph{inferring} that sources are located there, with the latter posing more stringent requirements. To this end, we present two figures of merit, which we refer to as ``$z$-$z$ plot'' and ``inference horizon'', that quantify the largest redshift one can possibly claim a source to be beyond. We argue that such inference requirements, in addition to detection requirements, should be investigated when quantifying the scientific payoff of future gravitational-wave facilities.
\end{abstract}

\maketitle

\section{Introduction}

Gravitational-wave (GW) interferometers LIGO and Virgo have detected about 90 signals from compact binary coalescences during their first three observing runs~\cite{LIGOScientific:2018mvr,LIGOScientific:2020ibl,LIGOScientific:2021djp}. The number of detections delivered by current facilities is expected to substantially increase in coming years~\cite{KAGRA:2013rdx,Baibhav:2019gxm}. At the same time, the GW community is actively preparing for the next major leap forward. Third-generation (3G) ground-based interferometers, namely the Einstein Telescope (ET) in Europe~\cite{Punturo:2010zz,Maggiore:2019uih} %
and Cosmic Explorer (CE) in the USA~\cite{Reitze:2019iox, Evans:2021gyd}, are now in advanced stage of planning and are featured on major strategic roadmaps~\cite{NAP26141,ESFRI}. Such future detectors will provide a GW-strain sensitivity that is about one order of magnitude higher compared to LIGO/Virgo, together with a broader detection band at both low and high frequencies. For compact binary mergers, this results in a $\mathcal{O}(10^2-10^4)$ %
increase in the number of detections, some of them at signal-to-noise ratio (SNR) as high as $\ssim 1000$ %
(for an overview of the science case of 3G detectors see Refs.~\cite{Maggiore:2019uih,Kalogera:2021bya}). 

In this context, one of the most interesting prospects is the possibility of observing GW signals from cosmological distances, signaling the presence of BHs at redshifts $z\gtrsim 2$, beyond the peak of the star-formation rate~\cite{Madau:2014bja,Vitale:2018yhm}. Even with future upgrades, current-generation interferometers are not expected to detect signals from redshifts higher than $z\lesssim 2$, while ET and CE could potentially reach $z\ssim100$ \cite{Ng:2020qpk, Iacovelli:2022bbs}. %
At these extreme distances, the possible populations of compact objects are greatly uncertain, hence providing open territory for discovery. After the peak of the star-formation rate, population III 
stars
made of metal-free gas originating from cosmological nucleosynthesis~\cite{Haemmerle:2020iqg,Klessen:2023qmc}
could lead to merging compact objects with distinguishable signatures~\cite{Kinugawa:2014zha, Hartwig:2016nde,Belczynski:2016ieo, Tanikawa:2020cca,Liu:2020lmi}. Binaries of primordial BHs formed from quantum fluctuations following the inflation~\cite{Garcia-Bellido:2017fdg,Sasaki:2018dmp}  are predicted to provide a merger-rate density that increases monotonically with redshift~\cite{Ali-Haimoud:2017rtz,Raidal:2018bbj,DeLuca:2020qqa,Franciolini:2021xbq}. 
The nature of the seeds of supermassive BHs observed at the center of galaxies is still uncertain~\cite{Volonteri:2021sfo} and might be elucidated by mass measurements of high-redshift GW sources.
In all cases, putative detections of BHs from such cosmological distances will be revolutionary for our (astro)physical understanding of the early Universe.

But, crucially, \emph{detecting} a signal originating from a high-redshift source does not imply \emph{inferring} that the source is   at that redshift. The requirements (in terms of both instrumental noise and data-analysis pipelines) for the latter are stronger compared to the former. Indeed, some recent studies \cite{Vitale:2016icu, Vitale:2018nif, Ng:2021sqn, Ng:2022vbz} indicate that, although 3G detectors will detect most, if not all, stellar-mass BH binaries in the Universe, they are unlikely to provide information on the distance to the source within $\lesssim 10\%$ whenever $z\gtrsim10$. %
The appropriate figure-of-merit  \cite{Hall:2019xmm} to forecast the cosmology science case of 3G detectors is, therefore, not whether the interferometers will {detect} GWs from high-redshift sources, but rather whether we will be able to confidently tell that the source is located where conventional BHs of stellar origin are not expected to be present.

In this paper, we present updated estimates on the redshift measurement errors expected for 3G detectors. We make use of the \textsc{gwfast} numerical pipeline~\cite{Iacovelli:2022bbs, Iacovelli:2022mbg} --a Fisher-matrix~\cite{Cutler:1994ys,Vallisneri:2007ev,Rodriguez:2013mla} code that was specifically designed for 3G forecasts and includes high-accuracy computation via automatic differentiation (cf. Refs.~\cite{Dupletsa:2022scg,Borhanian:2020ypi} for similar packages). In particular, we point out that a useful figure of merit for the cosmological case of ET and CE is provided by what we refer to as ``$z$-$z$ plots'', namely the lower {credible} interval on the redshift as a function of the source redshift itself. This is explored for both individual sources (Sec.~\ref{singlesource}) and broad populations (Sec.~\ref{broader}). We then show how that forecasts on the redshift constraints can be conveniently summarized in terms of an ``inference horizon'' for a specific detector configuration, to be compared against the common notion of ``detection horizon'' that expresses the source detectability (Sec.~\ref{horizon}). We conclude with a quick summary of our findings (Sec.~\ref{concl}). %
We use natural units where $c=G=1$.

\section{Redshift-redshift plots}
\label{singlesource}

We use the \textsc{gwfast} Fisher-matrix code as described in Refs.~\cite{Iacovelli:2022bbs, Iacovelli:2022mbg}. For each source, we simulate GW emission using the $\textsc{IMRPhenomHM}$ \cite{London:2017bcn} waveform model.  We consider a network of 3G detectors composed of ET in its triangular design located in Sardinia, %
and two CEs, one with 40-km long arms and one with 20-km long arms, located in Idaho and New Mexico respectively; see Ref.~\cite{Iacovelli:2022mbg} for details on the adopted noise curves and orientations. As standard practice in the field, we consider the SNR as a detection statistics\footnote{More precisely, we consider the amplitude SNR extracted from a perfect matched-filtering pipeline.} and adopt a threshold of $8$. %
For detectable sources, we compute the forecasted errors from the Fisher matrix $\Gamma_{ij}$.%

For BBHs with spins aligned along the axis $z$, the waveform parameters used in \textsc{gwfast} are %
\begin{equation}
\label{thetavector}
\theta = \{{\cal M}_c, \eta, \chi_{1,z}, \chi_{2,z}, d_L, \alpha, \delta, \iota, \psi, t_c, \Phi_c\}\,,    
\end{equation}
where ${\cal M}_c$ is the detector-frame chirp mass, $\eta$ is the symmetric mass ratio, $\chi_{i,z}$ are the dimensionless spins of the two BHs projected along the  orbital angular momentum of the binary, $d_L$ is the luminosity distance, $\alpha$ and $\delta$ are the right ascension and declination, respectively, %
$\iota$ is the inclination angle, $\psi$ is the polarization angle, $t_c$ is the time of coalescence, and $\Phi_c$ is the phase at coalescence.

The Fisher-matrix formalism consists of assuming that the likelihood %
for the parameters in Eq.~(\ref{thetavector}) is approximated by a multivariate Gaussian 
 with covariance $\Gamma^{-1}_{ij}$. %
In particular, the marginal likelihood for the {measured} luminosity distance {$d_{L, \rm obs}$} {given the true luminosity distance $d_{L,{\rm true}}$} is approximated by a Gaussian distribution $\mathcal{N}( d_{L, \rm obs}; d_{L,{\rm true}} ,  \sigma_{d_L} )$ where  $\sigma_{d_L} = \sqrt{\Gamma^{-1}_{d_L d_L}}$ (in practice,  this is also truncated to impose $d_L\geq 0$). %
For this work, we are rather interested in the posterior distribution on the redshift, marginalized on the other waveform parameters. For simplicity, we assume that the cosmology is known and consider a flat $\Lambda$CDM model with parameters from Ref.~\cite{Planck:2018vyg}. In this case, the redshift can be directly inferred from the luminosity distance $d_L$ measured in GWs by inverting the relation 
\begin{equation}\label{eq:dLofz}
    d_L(z) = \,(1+z) \int_{0}^z \dfrac{\dd{z'}}{H(z')}\,,
\end{equation}
where  $H(z)$ is the Hubble parameter \cite{Hogg:1999ad}. We can thus trade the luminosity distance  $d_L$ for the redshift $z$ in the  parameter vector $\theta$ and denote the remaining waveform parameters by $\bar \theta$ (i.e. $\theta= \{\bar\theta, z \}$) %

{Assuming a redshift prior that is uniform in comoving volume $V_c$ and source-frame  time \cite{Dominik:2014yma}} %
{one can write the redshift posterior as}
\begin{align}
\label{zpost}
p\left( z | z_{\rm true}, \bar \theta\right)
& \propto \mathcal{N}\left(d_{L,{\rm obs}}, ; d_L(z),  \sigma_{d_L} \right) \times \frac{\dd V_c}{\dd z} \frac{1}{1+z}\,, 
\end{align}
{where $z_{\rm true}$ is the true value of the redshift.\footnote{For all other parameters there will be no ambiguity between true and measured quantities, so the symbol  $\bar \theta$ simply refers to the true values.} We will assume an unbiased measurement and set $d_{L,{\rm obs}} = d_{L,{\rm true}}$}.
{The impact of the Bayesian prior on BH redshift measurements with 3G detectors has been explored in Ref.~\cite{Ng:2021sqn}}.

The goal of this paper is to investigate the inferred value $z_{c\%}(z_{\rm true}, \bar \theta)$ that provide a lower limit on our redshift estimate. That is: given our measurement, we can claim that the source is located at $z>z_{c\%}$ at $c\%$ credible interval. In symbols, this is
\begin{equation}\label{eq:Pzz}
P \left(  z \geq z_{c_{ \%}} | z_{\rm true},  {\bar \theta}  \right) = \int_{z_{c\%}
} ^\infty p \left(  z | z_{\rm true},  \bar \theta  \right) \dd z = \frac{c}{100}\,,
\end{equation}
where $p$ is the posterior probability density function and $P$ is the posterior cumulative distribution function. %
As common practice in GW astronomy, in the following we will often use $c=90$. 

\begin{figure}[t]
    \centering
    \includegraphics[width=.47\textwidth]{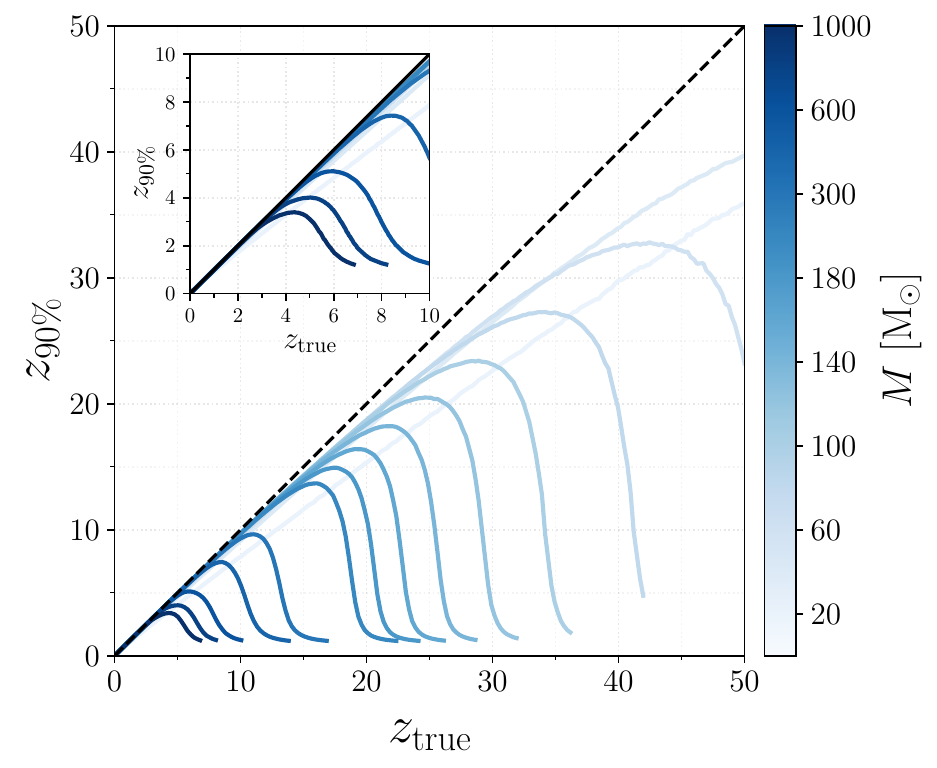}
    \caption{Lower bound on the estimated redshift  $z_{90\%}$ as a function of the true source redshift $z_{\rm true}$. We consider fiducial sources with different source-frame mass as reported on the color scale. 
    }
    \label{fig:zzplotSingleSource}
\end{figure}

Figure~\ref{fig:zzplotSingleSource} illustrates the behavior of  $z_{90\%}(z_{\rm true}, \bar \theta)$ as a function of $z_{\rm true}$. For this exercise, we consider a series of fiducial sources with fixed values of $\bar \theta$. In particular, we study binaries with total mass $M\in [20~{\rm M}_\odot,1000 ~{\rm M}_\odot]$ %
equal masses $\eta=0.25$, zero spins $\chi_{1,z}=\chi_{2,z}=0$, close-to-face-on orientation $\iota = 0.1$ (this is to avoid the ill-conditioning of the Fisher matrix for $\iota=0$), sky position $\alpha=\delta=\pi/4$, polarization angle $\psi=\pi/4$, GPS time $t_c$ and phase of  coalescence  $\Phi_c=0$. %

For a given source-frame total mass $M$,  $z_{90\%}$ reaches a maximum value at a redshift $z_{\rm peak}$ above which the confidence quickly degrades before the source becomes undetectable (this happens where the curves in Fig.~\ref{fig:zzplotSingleSource} terminate). %
This behavior can be understood considering the signal morphology in band, as both the amplitude and the frequency of GW signals decrease as the redshift increases {(cf. Ref.~\cite{Ng:2021sqn})}. %
While at first the value of $z_{90\%}$ remains monotonic with $z_{\rm true}$ (i.e. the increased error on the redshift is compensated by a larger median), sources at redshifts $\gtrsim z_{\rm peak}$ present fewer and fewer cycles in band such that the uncertainty on the luminosity distance increases enough to result in smaller values of $z_{90\%}$ (even though the median also increases).

\begin{figure}[t]
    \centering
    \includegraphics[width=\columnwidth]{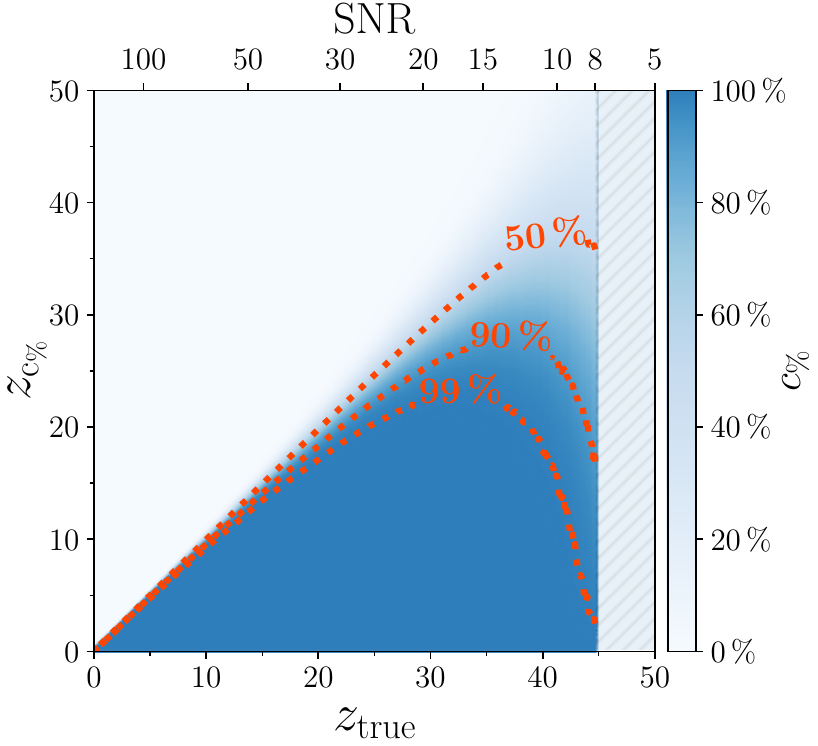}
    \caption{ $z$-$z$ plot for a GW150914-like source observed with a network of one ET and two CE instruments. A source at redshift $z_{\rm true}$ (bottom $x$-axis) can be placed at redshift larger than $z_{\rm c\%}$ ($y$-axis) at $c\%$ credible interval (color bar). Three representative  values $c = 50, 90,$ and $99$ are indicated with dashed contours. The source SNR is reported on the top $x$-axis. The hatched region to the right corresponds to sources below the detectability threshold of $\rm SNR=8$. %
    }   
    \label{fig:zzplotGW150914}
\end{figure}

Next, for a given source we study the probability  $ P \left(  z \geq z_{c_{ \%}} | z_{\rm true},  {\bar \theta}  \right) $ defined in Eq.~(\ref{eq:Pzz}) as a function of $z_{\rm true}$ and $z_{c_{ \%}}$.
This tells us, for each true value of the source redshift, the confidence with which we will be able to place a source beyond at a certain credibility. 
In analogy with the $p$-$p$ plot commonly used in statistics~\cite{gnanadesikan1968probability}, we refer to this visualization as ``$z$-$z$ plot.'' %
In Figure \ref{fig:zzplotGW150914} we show the $z$-$z$ plot of a GW150914-like source%
~\cite{LIGOScientific:2016aoc}. %
For instance, considering the $99\%$ contour, %
one infers that a source located at $z_{\rm true}\simeq 35$ can only be placed at redshift $z> z_{99\%}\simeq 23$ at $99\%$ credibility.
{Note how the $c=50$ contour if Fig.~\ref{fig:zzplotGW150914} does not lie long the plot diagonal, i.e. $z_{50\%} \lesssim z_{\rm true}$. This is because of the reduced prior volume at high redshifts, cf. Eq.(\ref{zpost}).} %

\begin{figure*}

         \includegraphics[width=0.97\columnwidth]{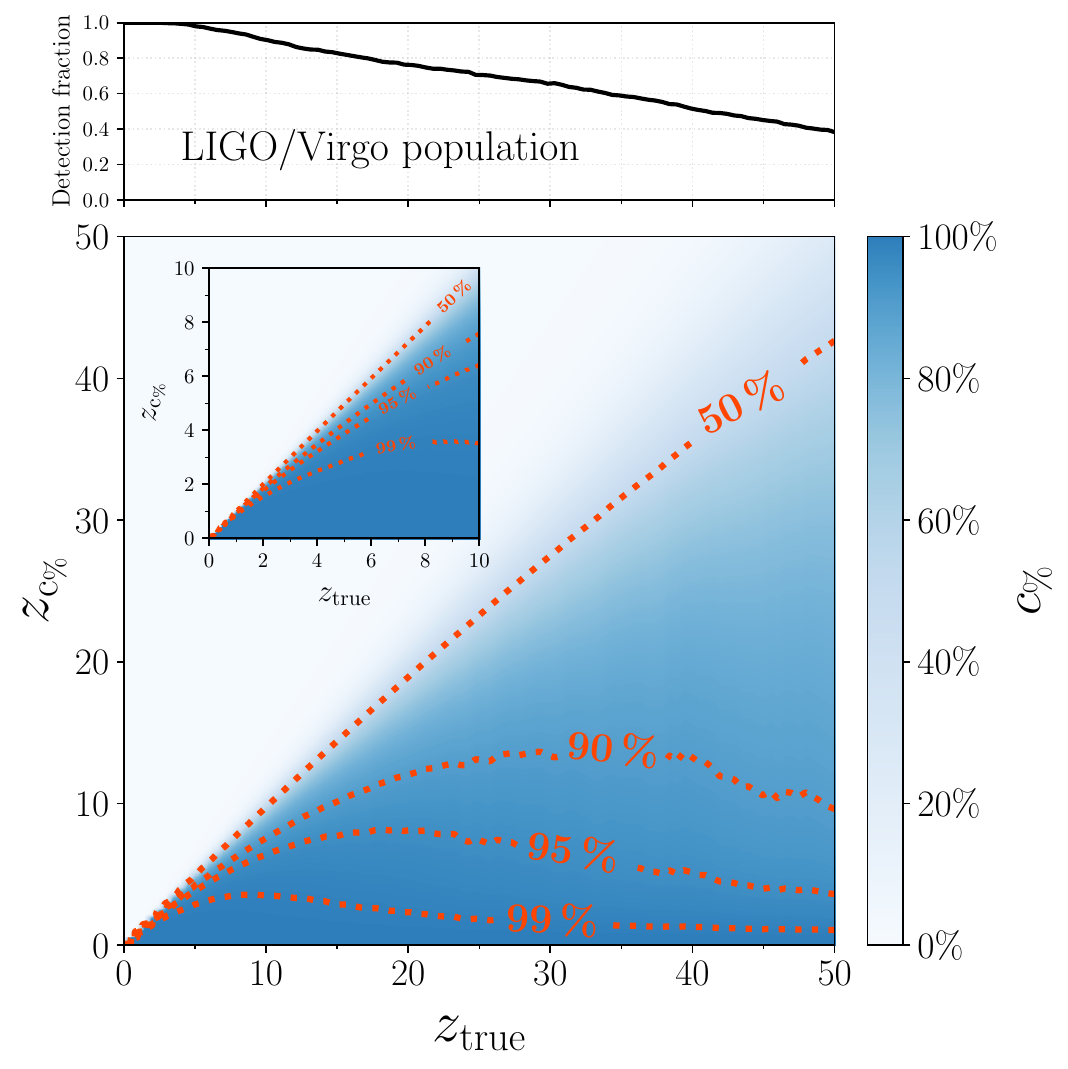}
        \hfill
        \includegraphics[width=0.97\columnwidth]{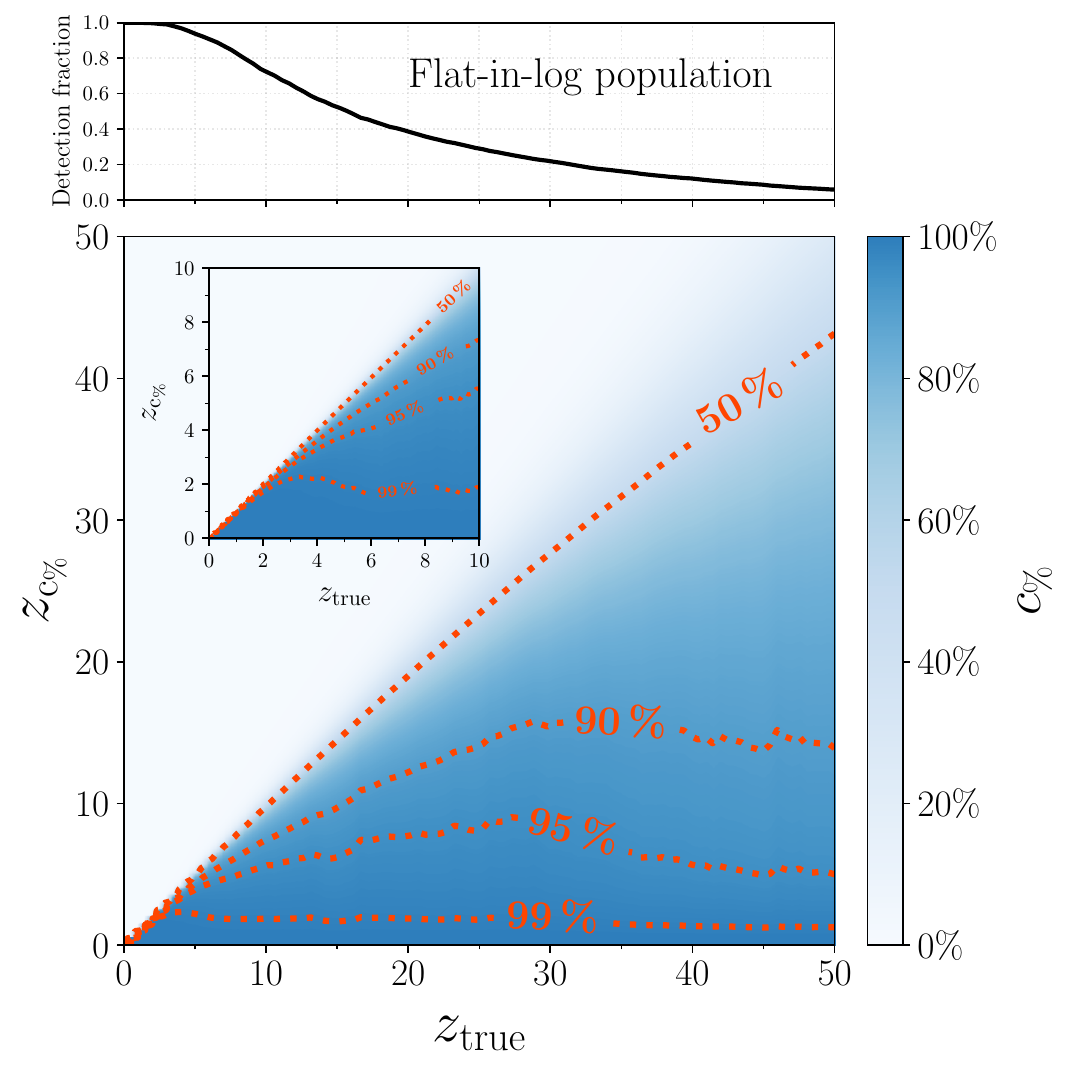}

\caption{Population-averaged $z$-$z$ plots (bottom panels) for two populations of binary BH population, one where we extrapolate current LIGO/Virgo results (left panel) and one where we assume a BH-mass spectrum that is flat in log-space between $5~{\rm M_\odot}$ and $500~{\rm M_ \odot}$ (right panel). We consider a network of ET and two CE instruments; the resulting fractions of detected sources are reported in the top panels.%
}
\label{fig:zzplotLVLpop}

\end{figure*}

\section{Broader populations}
\label{broader}

The $z$-$z$ plot presented in the previous section refers to individual BH binaries as it depends on the source parameters ${\bar \theta}$. Assuming sources are distributed according to a population   $p_{\rm pop}\left( {\bar \theta}  |  {\lambda} \right)$ with (hyper)parameters $\lambda$, one can also define a population-averaged analog of Eq.~(\ref{eq:Pzz}) as follows 
\begin{equation}\label{eq:PzzPop}
 P\! \left(  z \!\geq\! z_{c_{ \%}} | z_{\rm true},  \lambda  \right) \!\equiv\! \frac{ \displaystyle \int \!\! \dd{{\bar \theta}} \,P\!\left(  z \!\geq\! z_{c_{ \%}} | \theta  \right)  p_{\rm pop}\!\left( {\bar \theta}  |  {\lambda} \right)  P_{\rm det}(\theta)   }{ \displaystyle  \int\!\!  \dd{{\bar \theta}}\,  p_{\rm pop}\!\left( {\bar \theta}  |  {\lambda} \right)  P_{\rm det}(\theta) } ,
\end{equation}
where once again $\theta=\{ \bar\theta, z_{\rm true}\}$. The detection probability $P_{\rm det}(\theta)$ is, for our simple investigation, an indicator function that is unity if the SNR is $\geq8$ and zero otherwise. 

The expected distribution of BH binaries at high redshifts is highly uncertain --and thus needs to be discovered! Figure~\ref{fig:zzplotLVLpop} illustrates two examples assuming populations that are equally unrealistic. In the left panel, we extrapolate current LIGO/Virgo results at low redshifts. We generate binaries from the joint mass-spin distribution found in the population-inference analysis of Ref.~\cite{LIGOScientific:2021psn} adopting their \textsc{power-law+peak} and \textsc{default spin} phenomenological model and taking  for simplicity the maximum-a-posteriori value of each marginalized population-parameter distributions. In the right panel, we assume that the BH component masses are distributed uniformly in log in $[5~{\rm M}_\odot, 500~{\rm M}_\odot]$ and the aligned components of the spins are distributed  uniformly in $[-1,1]$. We place each of these two populations $p_{\rm pop}\left( {\bar \theta}  |  {\lambda} \right)$  at different (and fixed) redshifts $z_{\rm true}$ and compute the resulting averaged probability from Eq.~(\ref{eq:PzzPop}).

Although the two adopted populations are broadly different, they result in $z$-$z$ plots that are qualitatively similar (Fig.~\ref{fig:zzplotLVLpop}). The main difference between the assumptions is the high-mass cutoff, which is $< 100 {\rm M}_\odot$ for the sources extrapolated from LIGO/Virgo data and set to $500 {\rm M}_\odot$ for our simplistic flat-in-log distribution. This causes a difference in the  number of detected events (see the top panels in Fig.~\ref{fig:zzplotLVLpop}) but, once averaged over the \emph{detected} populations [cf. Eq.~(\ref{eq:PzzPop}) which contains $P_{\rm det}(\theta)$], the resulting expectations are largely unaffected. We interpret this is a relatively solid finding that takes into account the wide uncertainties on the expected population, at least within the limit of our assumptions. For instance, we find that, of all events detected {at redshift $\geq 30$}, only {$\ssim 3\%$ (21\%)} of them can be constrained to be above redshift {30} at 99\% (90\%) credibility {for the LIGO/Virgo population, while we find {$\ssim 3\%$ (15\%)} for the flat-in-log population}.

\section{Inference horizon}
\label{horizon}

\begin{figure}[t]
    \centering
    \includegraphics[width=\columnwidth]{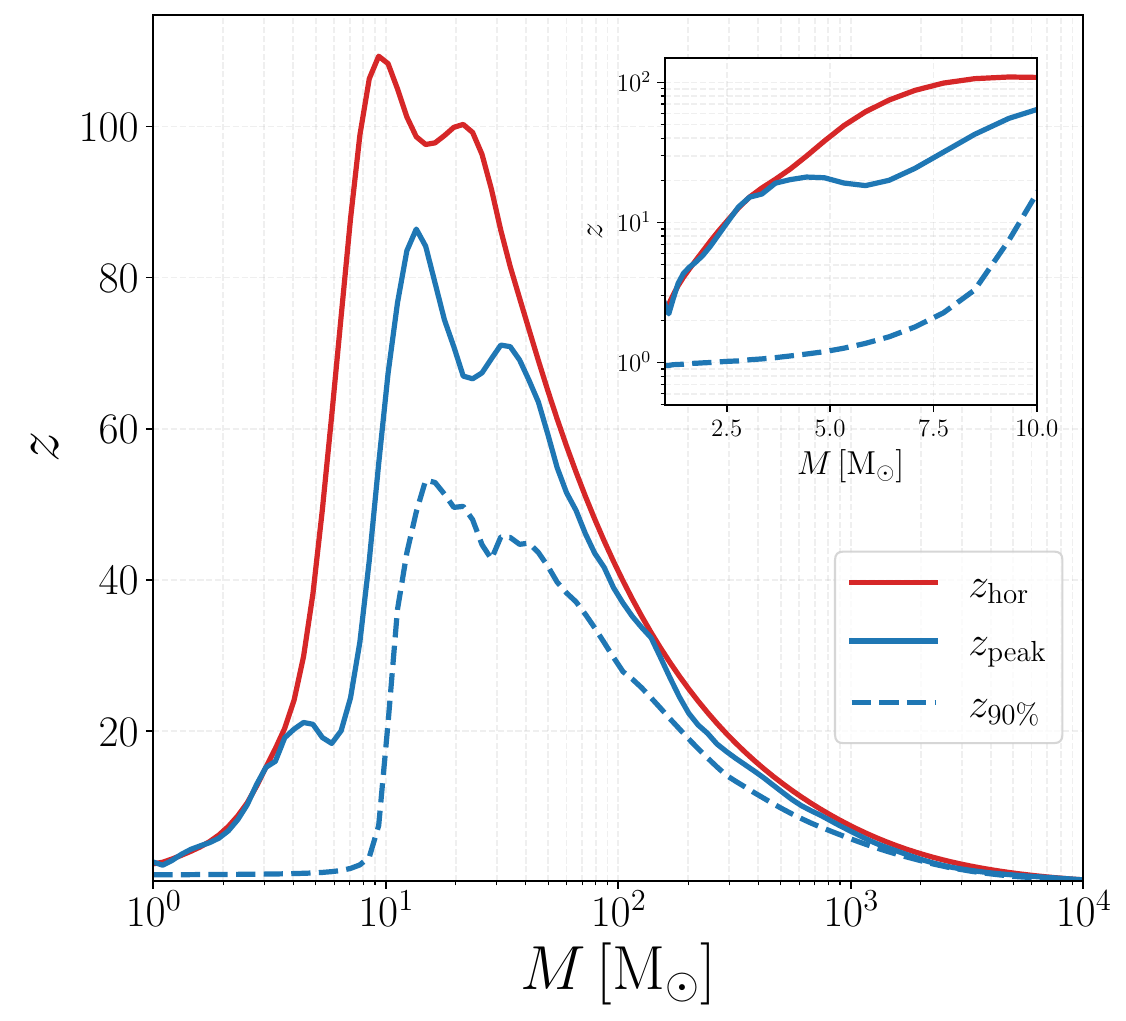}
    \caption{Detection and inference horizons for a 3G detector network made of a triangular ET and two L-shaped CE instruments. The red curve indicates the larger redshift at which a source of a given mass $M$ is detectable, $z_{\rm hor}$, the solid blue curve indicates the true value of the redshift $z_{\rm peak}$ that provides the most stringent constraint, and the dashed blue curves indicates the corresponding lower bound $z_{\rm 90\%}$ on the redshift itself. %
    }
    \label{fig:horizonNoLog}
\end{figure}
The notion of $z_{c\%}$ allows to introduce a further figure of merit for the reach of the detector. 
This is often framed in terms of the detector horizon, i.e. the maximum redshift to which a source with optimal orientation and sky location can be detected. %
But, as argued in this paper, a detection does not directly translate into added scientific potential if the resulting measurement is not sufficiently accurate. In this context, it is meaningful to define a notion of horizon for the confident placement of a source above a given redshift.%

Two of such ``inference horizon'' metrics are reported in  Fig.~\ref{fig:horizonNoLog}  for a network made of a triangular ET and two CE instruments, and assuming BHs with equal masses and zero spins. In particular we show: %
\begin{itemize}
\item[(i)] The usual detection horizon $z_{\rm hor}$, i.e., the maximum redshift to which an source with optimal inclination and sky location can be detected. While identifying the optimal source is trivial for a single detector~\cite{Schutz:2011tw}, considering a network requires a numerical maximization of the SNR. 
\item[(ii)] The value $z_{\rm peak}$ corresponding to the true value of the redshift where the $z_{90\%}$ is largest. This corresponds to the value reported on the $x$-axis for the maximum in curves likes those in Fig.~\ref{fig:zzplotSingleSource} (though unlike in Sec.~\ref{singlesource} here we also optimize over the extrinsic parameters).%
\item[(iii)] The value of $z_{90\%}$ at $z_{\rm peak}$. This is the largest redshift we can possible put a claim on for a given $M$ and corresponds to the $y$-axis value of the maximum in curves likes those in Fig~\ref{fig:zzplotSingleSource}. 
\end{itemize}

Figure~\ref{fig:horizonNoLog} should therefore be interpreted as follows.
Consider a given value of the total mass, say $M=20\,{\rm M}_{\odot}$. An optimal source with this mass would be detectable up to redshift $z_{\rm hor}\ssim 100$.  However, the source that provides the most stringent constraint on the redshift is located at {$z_{\rm peak}\ssim 70$}. %
But even for this lucky source, we only would  be able to infer that the redshift is larger than {$z_{90\%} \ssim 50$} at $90\%$ credibility.
Rephrasing the same argument, if one considers $z_{90\%}=30$ as a threshold value to indicate a primordial origin \cite{Koushiappas:2017kqm}, Fig.~\ref{fig:horizonNoLog} shows that such a potential claim can only be made for binaries of $\sim10-100\, M_{\odot}$, which is a narrower range compared to what indicated by the detector horizon (see also \cite{Franciolini:2023opt} for similar conclusions).

The double peak appearing in the curves of Fig.~\ref{fig:horizonNoLog} is due to the difference in sensitivity between ET and CE, with all three horizons being dominated by the most sensitive detector in the network (cf. Ref.~\cite{Iacovelli:2022bbs}).

\section{Conclusions}
\label{concl}

The characterization of sources at high redshifts --not just their detection-- is a key scientific target of 3G GW interferometers. With both the ET and CE projects in their advanced stage of planning, it is important to optimize their configurations using figures of merit that closely mirror the targeted science. In the context of high-redshift BH binaries, the most tantalizing possibility is that of confidently ruling out formation channels based on conventional stellar evolution, thus leaving room for sources of cosmological origin. %

We formulate this problem in terms of two quantifiers:
\begin{itemize}
\item[(i)]
In analogy with the so-called $p$-$p$ plot used in statistics, we argue that a ``$z$-$z$ plot'' indicating redshift bounds as a function of the redshift itself conveys the most relevant information in a compact fashion. 
\item[(ii)] Similarly, comparing an ``inference horizon'' (i.e. the largest redshift one can confidently put a bound on) to the usual ``detection horizon'' (i.e. the largest redshift one can detect a source from) provides a useful indicator of how the combined effect of detector sensitivity and data-analysis efficiency can be translated into constraints of (astro)physical nature.
\end{itemize}

\citeauthor{Ng:2021sqn}~\cite{Ng:2021sqn} performed Bayesian parameter estimation on a limited number of high-redshift sources observed with 3G detectors and concluded that primordial-origin claims based on single-events would be challenging; in our notation, their threshold for a ``safe'' inference is $z_{97\%}> 30$. Our results provide a framework to generalize those claims taking into account both the redshift dependence on the threshold and the redshift threshold itself. The assumed redshift prior might play a prominent role~\cite{Ng:2021sqn}; possible refinements in this direction include adopting population-informed priors~\cite{Moore:2021xhn}.

Overall, we find that confidently ruling out conventional star formation for single events might turn out to be very challenging, %
even with 3G detectors. More promising avenues to constrain BH binary formation at high redshifts include population analyses of resolved sources~\cite{Ng:2022agi} {(though handling model systematics will also be challenging)} and the characterization of the stochastic GW background~\cite{Mukherjee:2021itf,Bavera:2021wmw}.
We hope our investigation will provide additional tools to better quantify the performance of those advanced machines in view of their commissioning.

\acknowledgements
We thank Monica Colpi, Ken Ng, Matthew Mould, and Gabriele Franciolini for discussions. \textsc{gwfast} is publicly available at \href{https://github.com/cosmostatgw/gwfast}{github.com/cosmostatgw/gwfast}.
M.M. and D.G. are supported by European Union's H2020 ERC Starting Grant No.~945155--GWmining, Cariplo Foundation Grant No.~2021-0555, and the ICSC National Research Centre funded by NextGenerationEU. F.I. is supported by Swiss National Science Foundation Grant No.~200020-191957, the SwissMap National Center for Competence in Research, and the Istituto Svizzero ``Milano Calling'' fellowship.  D.G. is supported by Leverhulme Trust Grant No.~RPG-2019-350. Computational work was performed at CINECA with allocations through INFN, Bicocca, and ISCRA project HP10BEQ9JB.
This paper has ET report number No.~ET-0086A-23.

\bibliography{zzplot}

\begin{thebibliography}{55}%
\makeatletter
\providecommand \@ifxundefined [1]{%
 \@ifx{#1\undefined}
}%
\providecommand \@ifnum [1]{%
 \ifnum #1\expandafter \@firstoftwo
 \else \expandafter \@secondoftwo
 \fi
}%
\providecommand \@ifx [1]{%
 \ifx #1\expandafter \@firstoftwo
 \else \expandafter \@secondoftwo
 \fi
}%
\providecommand \natexlab [1]{#1}%
\providecommand \enquote  [1]{``#1''}%
\providecommand \bibnamefont  [1]{#1}%
\providecommand \bibfnamefont [1]{#1}%
\providecommand \citenamefont [1]{#1}%
\providecommand \href@noop [0]{\@secondoftwo}%
\providecommand \href [0]{\begingroup \@sanitize@url \@href}%
\providecommand \@href[1]{\@@startlink{#1}\@@href}%
\providecommand \@@href[1]{\endgroup#1\@@endlink}%
\providecommand \@sanitize@url [0]{\catcode `\\12\catcode `\$12\catcode
  `\&12\catcode `\#12\catcode `\^12\catcode `\_12\catcode `\%12\relax}%
\providecommand \@@startlink[1]{}%
\providecommand \@@endlink[0]{}%
\providecommand \url  [0]{\begingroup\@sanitize@url \@url }%
\providecommand \@url [1]{\endgroup\@href {#1}{\urlprefix }}%
\providecommand \urlprefix  [0]{URL }%
\providecommand \Eprint [0]{\href }%
\providecommand \doibase [0]{http://dx.doi.org/}%
\providecommand \selectlanguage [0]{\@gobble}%
\providecommand \bibinfo  [0]{\@secondoftwo}%
\providecommand \bibfield  [0]{\@secondoftwo}%
\providecommand \translation [1]{[#1]}%
\providecommand \BibitemOpen [0]{}%
\providecommand \bibitemStop [0]{}%
\providecommand \bibitemNoStop [0]{.\EOS\space}%
\providecommand \EOS [0]{\spacefactor3000\relax}%
\providecommand \BibitemShut  [1]{\csname bibitem#1\endcsname}%
\let\auto@bib@innerbib\@empty
\bibitem [{\citenamefont {Abbott}\ \emph {et~al.}(2019)\citenamefont {Abbott}
  \emph {et~al.}}]{LIGOScientific:2018mvr}%
  \BibitemOpen
  \bibfield  {author} {\bibinfo {author} {\bibfnamefont {B.~P.}\ \bibnamefont
  {Abbott}} \emph {et~al.} (\bibinfo {collaboration} {LIGO Scientific,
  Virgo}),\ }\href {\doibase 10.1103/PhysRevX.9.031040} {\bibfield  {journal}
  {\bibinfo  {journal} {Phys. Rev. X}\ }\textbf {\bibinfo {volume} {9}},\
  \bibinfo {pages} {031040} (\bibinfo {year} {2019})},\ \Eprint
  {http://arxiv.org/abs/1811.12907} {arXiv:1811.12907 [astro-ph.HE]}
  \BibitemShut {NoStop}%
\bibitem [{\citenamefont {Abbott}\ \emph
  {et~al.}(2021{\natexlab{a}})\citenamefont {Abbott} \emph
  {et~al.}}]{LIGOScientific:2020ibl}%
  \BibitemOpen
  \bibfield  {author} {\bibinfo {author} {\bibfnamefont {R.}~\bibnamefont
  {Abbott}} \emph {et~al.} (\bibinfo {collaboration} {LIGO Scientific,
  Virgo}),\ }\href {\doibase 10.1103/PhysRevX.11.021053} {\bibfield  {journal}
  {\bibinfo  {journal} {Phys. Rev. X}\ }\textbf {\bibinfo {volume} {11}},\
  \bibinfo {pages} {021053} (\bibinfo {year} {2021}{\natexlab{a}})},\ \Eprint
  {http://arxiv.org/abs/2010.14527} {arXiv:2010.14527 [gr-qc]} \BibitemShut
  {NoStop}%
\bibitem [{\citenamefont {Abbott}\ \emph
  {et~al.}(2021{\natexlab{b}})\citenamefont {Abbott} \emph
  {et~al.}}]{LIGOScientific:2021djp}%
  \BibitemOpen
  \bibfield  {author} {\bibinfo {author} {\bibfnamefont {R.}~\bibnamefont
  {Abbott}} \emph {et~al.} (\bibinfo {collaboration} {LIGO Scientific, Virgo,
  KAGRA}),\ }\href@noop {} {\  (\bibinfo {year} {2021}{\natexlab{b}})},\
  \Eprint {http://arxiv.org/abs/2111.03606} {arXiv:2111.03606 [gr-qc]}
  \BibitemShut {NoStop}%
\bibitem [{\citenamefont {Abbott}\ \emph {et~al.}(2018)\citenamefont {Abbott}
  \emph {et~al.}}]{KAGRA:2013rdx}%
  \BibitemOpen
  \bibfield  {author} {\bibinfo {author} {\bibfnamefont {B.~P.}\ \bibnamefont
  {Abbott}} \emph {et~al.} (\bibinfo {collaboration} {LIGO Scientific, Virgo,
  KAGRA}),\ }\href {\doibase 10.1007/s41114-020-00026-9} {\bibfield  {journal}
  {\bibinfo  {journal} {Living Rev. Rel.}\ }\textbf {\bibinfo {volume} {21}},\
  \bibinfo {pages} {3} (\bibinfo {year} {2018})},\ \Eprint
  {http://arxiv.org/abs/1304.0670} {arXiv:1304.0670 [gr-qc]} \BibitemShut
  {NoStop}%
\bibitem [{\citenamefont {Baibhav}\ \emph {et~al.}(2019)\citenamefont
  {Baibhav}, \citenamefont {Berti}, \citenamefont {Gerosa}, \citenamefont
  {Mapelli}, \citenamefont {Giacobbo}, \citenamefont {Bouffanais},\ and\
  \citenamefont {Di~Carlo}}]{Baibhav:2019gxm}%
  \BibitemOpen
  \bibfield  {author} {\bibinfo {author} {\bibfnamefont {V.}~\bibnamefont
  {Baibhav}}, \bibinfo {author} {\bibfnamefont {E.}~\bibnamefont {Berti}},
  \bibinfo {author} {\bibfnamefont {D.}~\bibnamefont {Gerosa}}, \bibinfo
  {author} {\bibfnamefont {M.}~\bibnamefont {Mapelli}}, \bibinfo {author}
  {\bibfnamefont {N.}~\bibnamefont {Giacobbo}}, \bibinfo {author}
  {\bibfnamefont {Y.}~\bibnamefont {Bouffanais}}, \ and\ \bibinfo {author}
  {\bibfnamefont {U.~N.}\ \bibnamefont {Di~Carlo}},\ }\href {\doibase
  10.1103/PhysRevD.100.064060} {\bibfield  {journal} {\bibinfo  {journal}
  {Phys. Rev. D}\ }\textbf {\bibinfo {volume} {100}},\ \bibinfo {pages}
  {064060} (\bibinfo {year} {2019})},\ \Eprint
  {http://arxiv.org/abs/1906.04197} {arXiv:1906.04197 [gr-qc]} \BibitemShut
  {NoStop}%
\bibitem [{\citenamefont {Punturo}\ \emph {et~al.}(2010)\citenamefont {Punturo}
  \emph {et~al.}}]{Punturo:2010zz}%
  \BibitemOpen
  \bibfield  {author} {\bibinfo {author} {\bibfnamefont {M.}~\bibnamefont
  {Punturo}} \emph {et~al.},\ }\href {\doibase 10.1088/0264-9381/27/19/194002}
  {\bibfield  {journal} {\bibinfo  {journal} {Class. Quant. Grav.}\ }\textbf
  {\bibinfo {volume} {27}},\ \bibinfo {pages} {194002} (\bibinfo {year}
  {2010})}\BibitemShut {NoStop}%
\bibitem [{\citenamefont {Maggiore}\ \emph {et~al.}(2020)\citenamefont
  {Maggiore} \emph {et~al.}}]{Maggiore:2019uih}%
  \BibitemOpen
  \bibfield  {author} {\bibinfo {author} {\bibfnamefont {M.}~\bibnamefont
  {Maggiore}} \emph {et~al.},\ }\href {\doibase 10.1088/1475-7516/2020/03/050}
  {\bibfield  {journal} {\bibinfo  {journal} {J. Cosmol. Astropart. Phys.}\
  }\textbf {\bibinfo {volume} {03}},\ \bibinfo {pages} {050} (\bibinfo {year}
  {2020})},\ \Eprint {http://arxiv.org/abs/1912.02622} {arXiv:1912.02622
  [astro-ph.CO]} \BibitemShut {NoStop}%
\bibitem [{\citenamefont {Reitze}\ \emph {et~al.}(2019)\citenamefont {Reitze}
  \emph {et~al.}}]{Reitze:2019iox}%
  \BibitemOpen
  \bibfield  {author} {\bibinfo {author} {\bibfnamefont {D.}~\bibnamefont
  {Reitze}} \emph {et~al.},\ }\href@noop {} {\bibfield  {journal} {\bibinfo
  {journal} {Bull. Am. Astron. Soc.}\ }\textbf {\bibinfo {volume} {51}},\
  \bibinfo {pages} {035} (\bibinfo {year} {2019})},\ \Eprint
  {http://arxiv.org/abs/1907.04833} {arXiv:1907.04833 [astro-ph.IM]}
  \BibitemShut {NoStop}%
\bibitem [{\citenamefont {Evans}\ \emph {et~al.}(2021)\citenamefont {Evans}
  \emph {et~al.}}]{Evans:2021gyd}%
  \BibitemOpen
  \bibfield  {author} {\bibinfo {author} {\bibfnamefont {M.}~\bibnamefont
  {Evans}} \emph {et~al.},\ }\href@noop {} {\  (\bibinfo {year} {2021})},\
  \Eprint {http://arxiv.org/abs/2109.09882} {arXiv:2109.09882 [astro-ph.IM]}
  \BibitemShut {NoStop}%
\bibitem [{\citenamefont {{National Academies of Sciences, Engineering, and
  Medicine}}(2021)}]{NAP26141}%
  \BibitemOpen
  \bibfield  {author} {\bibinfo {author} {\bibnamefont {{National Academies of
  Sciences, Engineering, and Medicine}}},\ }\href {\doibase 10.17226/26141}
  {\emph {\bibinfo {title} {Pathways to Discovery in Astronomy and Astrophysics
  for the 2020s}}}\ (\bibinfo {year} {2021})\BibitemShut {NoStop}%
\bibitem [{\citenamefont {{European Strategy Forum on Research
  Infrastructures}}(2021)}]{ESFRI}%
  \BibitemOpen
  \bibfield  {author} {\bibinfo {author} {\bibnamefont {{European Strategy
  Forum on Research Infrastructures}}},\ }\href {https://roadmap2021.esfri.eu/}
  {\emph {\bibinfo {title} {ESFRI Roadmap 2021}}}\ (\bibinfo {year}
  {2021})\BibitemShut {NoStop}%
\bibitem [{\citenamefont {Kalogera}\ \emph {et~al.}(2021)\citenamefont
  {Kalogera} \emph {et~al.}}]{Kalogera:2021bya}%
  \BibitemOpen
  \bibfield  {author} {\bibinfo {author} {\bibfnamefont {V.}~\bibnamefont
  {Kalogera}} \emph {et~al.},\ }\href@noop {} {\  (\bibinfo {year} {2021})},\
  \Eprint {http://arxiv.org/abs/2111.06990} {arXiv:2111.06990 [gr-qc]}
  \BibitemShut {NoStop}%
\bibitem [{\citenamefont {Madau}\ and\ \citenamefont
  {Dickinson}(2014)}]{Madau:2014bja}%
  \BibitemOpen
  \bibfield  {author} {\bibinfo {author} {\bibfnamefont {P.}~\bibnamefont
  {Madau}}\ and\ \bibinfo {author} {\bibfnamefont {M.}~\bibnamefont
  {Dickinson}},\ }\href {\doibase 10.1146/annurev-astro-081811-125615}
  {\bibfield  {journal} {\bibinfo  {journal} {Ann. Rev. Astron. Astrophys.}\
  }\textbf {\bibinfo {volume} {52}},\ \bibinfo {pages} {415} (\bibinfo {year}
  {2014})},\ \Eprint {http://arxiv.org/abs/1403.0007} {arXiv:1403.0007
  [astro-ph.CO]} \BibitemShut {NoStop}%
\bibitem [{\citenamefont {Vitale}\ \emph {et~al.}(2019)\citenamefont {Vitale},
  \citenamefont {Farr}, \citenamefont {Ng},\ and\ \citenamefont
  {Rodriguez}}]{Vitale:2018yhm}%
  \BibitemOpen
  \bibfield  {author} {\bibinfo {author} {\bibfnamefont {S.}~\bibnamefont
  {Vitale}}, \bibinfo {author} {\bibfnamefont {W.~M.}\ \bibnamefont {Farr}},
  \bibinfo {author} {\bibfnamefont {K.}~\bibnamefont {Ng}}, \ and\ \bibinfo
  {author} {\bibfnamefont {C.~L.}\ \bibnamefont {Rodriguez}},\ }\href {\doibase
  10.3847/2041-8213/ab50c0} {\bibfield  {journal} {\bibinfo  {journal}
  {Astrophys. J. Lett.}\ }\textbf {\bibinfo {volume} {886}},\ \bibinfo {pages}
  {L1} (\bibinfo {year} {2019})},\ \Eprint {http://arxiv.org/abs/1808.00901}
  {arXiv:1808.00901 [astro-ph.HE]} \BibitemShut {NoStop}%
\bibitem [{\citenamefont {Ng}\ \emph {et~al.}(2021)\citenamefont {Ng},
  \citenamefont {Vitale}, \citenamefont {Farr},\ and\ \citenamefont
  {Rodriguez}}]{Ng:2020qpk}%
  \BibitemOpen
  \bibfield  {author} {\bibinfo {author} {\bibfnamefont {K.~K.~Y.}\
  \bibnamefont {Ng}}, \bibinfo {author} {\bibfnamefont {S.}~\bibnamefont
  {Vitale}}, \bibinfo {author} {\bibfnamefont {W.~M.}\ \bibnamefont {Farr}}, \
  and\ \bibinfo {author} {\bibfnamefont {C.~L.}\ \bibnamefont {Rodriguez}},\
  }\href {\doibase 10.3847/2041-8213/abf8be} {\bibfield  {journal} {\bibinfo
  {journal} {Astrophys. J. Lett.}\ }\textbf {\bibinfo {volume} {913}},\
  \bibinfo {pages} {L5} (\bibinfo {year} {2021})},\ \Eprint
  {http://arxiv.org/abs/2012.09876} {arXiv:2012.09876 [astro-ph.CO]}
  \BibitemShut {NoStop}%
\bibitem [{\citenamefont {Iacovelli}\ \emph
  {et~al.}(2022{\natexlab{a}})\citenamefont {Iacovelli}, \citenamefont
  {Mancarella}, \citenamefont {Foffa},\ and\ \citenamefont
  {Maggiore}}]{Iacovelli:2022bbs}%
  \BibitemOpen
  \bibfield  {author} {\bibinfo {author} {\bibfnamefont {F.}~\bibnamefont
  {Iacovelli}}, \bibinfo {author} {\bibfnamefont {M.}~\bibnamefont
  {Mancarella}}, \bibinfo {author} {\bibfnamefont {S.}~\bibnamefont {Foffa}}, \
  and\ \bibinfo {author} {\bibfnamefont {M.}~\bibnamefont {Maggiore}},\ }\href
  {\doibase 10.3847/1538-4357/ac9cd4} {\bibfield  {journal} {\bibinfo
  {journal} {Astrophys. J.}\ }\textbf {\bibinfo {volume} {941}},\ \bibinfo
  {pages} {208} (\bibinfo {year} {2022}{\natexlab{a}})},\ \Eprint
  {http://arxiv.org/abs/2207.02771} {arXiv:2207.02771 [gr-qc]} \BibitemShut
  {NoStop}%
\bibitem [{\citenamefont {Haemmerl\'e}\ \emph {et~al.}(2020)\citenamefont
  {Haemmerl\'e}, \citenamefont {Mayer}, \citenamefont {Klessen}, \citenamefont
  {Hosokawa}, \citenamefont {Madau},\ and\ \citenamefont
  {Bromm}}]{Haemmerle:2020iqg}%
  \BibitemOpen
  \bibfield  {author} {\bibinfo {author} {\bibfnamefont {L.}~\bibnamefont
  {Haemmerl\'e}}, \bibinfo {author} {\bibfnamefont {L.}~\bibnamefont {Mayer}},
  \bibinfo {author} {\bibfnamefont {R.~S.}\ \bibnamefont {Klessen}}, \bibinfo
  {author} {\bibfnamefont {T.}~\bibnamefont {Hosokawa}}, \bibinfo {author}
  {\bibfnamefont {P.}~\bibnamefont {Madau}}, \ and\ \bibinfo {author}
  {\bibfnamefont {V.}~\bibnamefont {Bromm}},\ }\href {\doibase
  10.1007/s11214-020-00673-y} {\bibfield  {journal} {\bibinfo  {journal} {Space
  Sci. Rev.}\ }\textbf {\bibinfo {volume} {216}},\ \bibinfo {pages} {48}
  (\bibinfo {year} {2020})},\ \Eprint {http://arxiv.org/abs/2003.10533}
  {arXiv:2003.10533 [astro-ph.GA]} \BibitemShut {NoStop}%
\bibitem [{\citenamefont {Klessen}\ and\ \citenamefont
  {Glover}(2023)}]{Klessen:2023qmc}%
  \BibitemOpen
  \bibfield  {author} {\bibinfo {author} {\bibfnamefont {R.~S.}\ \bibnamefont
  {Klessen}}\ and\ \bibinfo {author} {\bibfnamefont {S.~C.~O.}\ \bibnamefont
  {Glover}},\ }\href@noop {} {\  (\bibinfo {year} {2023})},\ \Eprint
  {http://arxiv.org/abs/2303.12500} {arXiv:2303.12500 [astro-ph.CO]}
  \BibitemShut {NoStop}%
\bibitem [{\citenamefont {Kinugawa}\ \emph {et~al.}(2014)\citenamefont
  {Kinugawa}, \citenamefont {Inayoshi}, \citenamefont {Hotokezaka},
  \citenamefont {Nakauchi},\ and\ \citenamefont {Nakamura}}]{Kinugawa:2014zha}%
  \BibitemOpen
  \bibfield  {author} {\bibinfo {author} {\bibfnamefont {T.}~\bibnamefont
  {Kinugawa}}, \bibinfo {author} {\bibfnamefont {K.}~\bibnamefont {Inayoshi}},
  \bibinfo {author} {\bibfnamefont {K.}~\bibnamefont {Hotokezaka}}, \bibinfo
  {author} {\bibfnamefont {D.}~\bibnamefont {Nakauchi}}, \ and\ \bibinfo
  {author} {\bibfnamefont {T.}~\bibnamefont {Nakamura}},\ }\href {\doibase
  10.1093/mnras/stu1022} {\bibfield  {journal} {\bibinfo  {journal} {Mon. Not.
  Roy. Astron. Soc.}\ }\textbf {\bibinfo {volume} {442}},\ \bibinfo {pages}
  {2963} (\bibinfo {year} {2014})},\ \Eprint {http://arxiv.org/abs/1402.6672}
  {arXiv:1402.6672 [astro-ph.HE]} \BibitemShut {NoStop}%
\bibitem [{\citenamefont {Hartwig}\ \emph {et~al.}(2016)\citenamefont
  {Hartwig}, \citenamefont {Volonteri}, \citenamefont {Bromm}, \citenamefont
  {Klessen}, \citenamefont {Barausse}, \citenamefont {Magg},\ and\
  \citenamefont {Stacy}}]{Hartwig:2016nde}%
  \BibitemOpen
  \bibfield  {author} {\bibinfo {author} {\bibfnamefont {T.}~\bibnamefont
  {Hartwig}}, \bibinfo {author} {\bibfnamefont {M.}~\bibnamefont {Volonteri}},
  \bibinfo {author} {\bibfnamefont {V.}~\bibnamefont {Bromm}}, \bibinfo
  {author} {\bibfnamefont {R.~S.}\ \bibnamefont {Klessen}}, \bibinfo {author}
  {\bibfnamefont {E.}~\bibnamefont {Barausse}}, \bibinfo {author}
  {\bibfnamefont {M.}~\bibnamefont {Magg}}, \ and\ \bibinfo {author}
  {\bibfnamefont {A.}~\bibnamefont {Stacy}},\ }\href {\doibase
  10.1093/mnrasl/slw074} {\bibfield  {journal} {\bibinfo  {journal} {Mon. Not.
  Roy. Astron. Soc.}\ }\textbf {\bibinfo {volume} {460}},\ \bibinfo {pages}
  {L74} (\bibinfo {year} {2016})},\ \Eprint {http://arxiv.org/abs/1603.05655}
  {arXiv:1603.05655 [astro-ph.GA]} \BibitemShut {NoStop}%
\bibitem [{\citenamefont {Belczynski}\ \emph {et~al.}(2017)\citenamefont
  {Belczynski}, \citenamefont {Ryu}, \citenamefont {Perna}, \citenamefont
  {Berti}, \citenamefont {Tanaka},\ and\ \citenamefont
  {Bulik}}]{Belczynski:2016ieo}%
  \BibitemOpen
  \bibfield  {author} {\bibinfo {author} {\bibfnamefont {K.}~\bibnamefont
  {Belczynski}}, \bibinfo {author} {\bibfnamefont {T.}~\bibnamefont {Ryu}},
  \bibinfo {author} {\bibfnamefont {R.}~\bibnamefont {Perna}}, \bibinfo
  {author} {\bibfnamefont {E.}~\bibnamefont {Berti}}, \bibinfo {author}
  {\bibfnamefont {T.~L.}\ \bibnamefont {Tanaka}}, \ and\ \bibinfo {author}
  {\bibfnamefont {T.}~\bibnamefont {Bulik}},\ }\href {\doibase
  10.1093/mnras/stx1759} {\bibfield  {journal} {\bibinfo  {journal} {Mon. Not.
  Roy. Astron. Soc.}\ }\textbf {\bibinfo {volume} {471}},\ \bibinfo {pages}
  {4702} (\bibinfo {year} {2017})},\ \Eprint {http://arxiv.org/abs/1612.01524}
  {arXiv:1612.01524 [astro-ph.HE]} \BibitemShut {NoStop}%
\bibitem [{\citenamefont {Tanikawa}\ \emph {et~al.}(2021)\citenamefont
  {Tanikawa}, \citenamefont {Susa}, \citenamefont {Yoshida}, \citenamefont
  {Trani},\ and\ \citenamefont {Kinugawa}}]{Tanikawa:2020cca}%
  \BibitemOpen
  \bibfield  {author} {\bibinfo {author} {\bibfnamefont {A.}~\bibnamefont
  {Tanikawa}}, \bibinfo {author} {\bibfnamefont {H.}~\bibnamefont {Susa}},
  \bibinfo {author} {\bibfnamefont {T.}~\bibnamefont {Yoshida}}, \bibinfo
  {author} {\bibfnamefont {A.~A.}\ \bibnamefont {Trani}}, \ and\ \bibinfo
  {author} {\bibfnamefont {T.}~\bibnamefont {Kinugawa}},\ }\href {\doibase
  10.3847/1538-4357/abe40d} {\bibfield  {journal} {\bibinfo  {journal}
  {Astrophys. J.}\ }\textbf {\bibinfo {volume} {910}},\ \bibinfo {pages} {30}
  (\bibinfo {year} {2021})},\ \Eprint {http://arxiv.org/abs/2008.01890}
  {arXiv:2008.01890 [astro-ph.HE]} \BibitemShut {NoStop}%
\bibitem [{\citenamefont {Liu}\ and\ \citenamefont
  {Bromm}(2020)}]{Liu:2020lmi}%
  \BibitemOpen
  \bibfield  {author} {\bibinfo {author} {\bibfnamefont {B.}~\bibnamefont
  {Liu}}\ and\ \bibinfo {author} {\bibfnamefont {V.}~\bibnamefont {Bromm}},\
  }\href {\doibase 10.3847/2041-8213/abc552} {\bibfield  {journal} {\bibinfo
  {journal} {Astrophys. J. Lett.}\ }\textbf {\bibinfo {volume} {903}},\
  \bibinfo {pages} {L40} (\bibinfo {year} {2020})},\ \Eprint
  {http://arxiv.org/abs/2009.11447} {arXiv:2009.11447 [astro-ph.GA]}
  \BibitemShut {NoStop}%
\bibitem [{\citenamefont {Garc\'\i{}a-Bellido}(2017)}]{Garcia-Bellido:2017fdg}%
  \BibitemOpen
  \bibfield  {author} {\bibinfo {author} {\bibfnamefont {J.}~\bibnamefont
  {Garc\'\i{}a-Bellido}},\ }\href {\doibase 10.1088/1742-6596/840/1/012032}
  {\bibfield  {journal} {\bibinfo  {journal} {J. Phys. Conf. Ser.}\ }\textbf
  {\bibinfo {volume} {840}},\ \bibinfo {pages} {012032} (\bibinfo {year}
  {2017})},\ \Eprint {http://arxiv.org/abs/1702.08275} {arXiv:1702.08275
  [astro-ph.CO]} \BibitemShut {NoStop}%
\bibitem [{\citenamefont {Sasaki}\ \emph {et~al.}(2018)\citenamefont {Sasaki},
  \citenamefont {Suyama}, \citenamefont {Tanaka},\ and\ \citenamefont
  {Yokoyama}}]{Sasaki:2018dmp}%
  \BibitemOpen
  \bibfield  {author} {\bibinfo {author} {\bibfnamefont {M.}~\bibnamefont
  {Sasaki}}, \bibinfo {author} {\bibfnamefont {T.}~\bibnamefont {Suyama}},
  \bibinfo {author} {\bibfnamefont {T.}~\bibnamefont {Tanaka}}, \ and\ \bibinfo
  {author} {\bibfnamefont {S.}~\bibnamefont {Yokoyama}},\ }\href {\doibase
  10.1088/1361-6382/aaa7b4} {\bibfield  {journal} {\bibinfo  {journal} {Class.
  Quant. Grav.}\ }\textbf {\bibinfo {volume} {35}},\ \bibinfo {pages} {063001}
  (\bibinfo {year} {2018})},\ \Eprint {http://arxiv.org/abs/1801.05235}
  {arXiv:1801.05235 [astro-ph.CO]} \BibitemShut {NoStop}%
\bibitem [{\citenamefont {Ali-Ha\"\i{}moud}\ \emph {et~al.}(2017)\citenamefont
  {Ali-Ha\"\i{}moud}, \citenamefont {Kovetz},\ and\ \citenamefont
  {Kamionkowski}}]{Ali-Haimoud:2017rtz}%
  \BibitemOpen
  \bibfield  {author} {\bibinfo {author} {\bibfnamefont {Y.}~\bibnamefont
  {Ali-Ha\"\i{}moud}}, \bibinfo {author} {\bibfnamefont {E.~D.}\ \bibnamefont
  {Kovetz}}, \ and\ \bibinfo {author} {\bibfnamefont {M.}~\bibnamefont
  {Kamionkowski}},\ }\href {\doibase 10.1103/PhysRevD.96.123523} {\bibfield
  {journal} {\bibinfo  {journal} {Phys. Rev. D}\ }\textbf {\bibinfo {volume}
  {96}},\ \bibinfo {pages} {123523} (\bibinfo {year} {2017})},\ \Eprint
  {http://arxiv.org/abs/1709.06576} {arXiv:1709.06576 [astro-ph.CO]}
  \BibitemShut {NoStop}%
\bibitem [{\citenamefont {Raidal}\ \emph {et~al.}(2019)\citenamefont {Raidal},
  \citenamefont {Spethmann}, \citenamefont {Vaskonen},\ and\ \citenamefont
  {Veerm\"ae}}]{Raidal:2018bbj}%
  \BibitemOpen
  \bibfield  {author} {\bibinfo {author} {\bibfnamefont {M.}~\bibnamefont
  {Raidal}}, \bibinfo {author} {\bibfnamefont {C.}~\bibnamefont {Spethmann}},
  \bibinfo {author} {\bibfnamefont {V.}~\bibnamefont {Vaskonen}}, \ and\
  \bibinfo {author} {\bibfnamefont {H.}~\bibnamefont {Veerm\"ae}},\ }\href
  {\doibase 10.1088/1475-7516/2019/02/018} {\bibfield  {journal} {\bibinfo
  {journal} {J. Cosmol. Astropart. Phys.}\ }\textbf {\bibinfo {volume} {02}},\
  \bibinfo {pages} {018} (\bibinfo {year} {2019})},\ \Eprint
  {http://arxiv.org/abs/1812.01930} {arXiv:1812.01930 [astro-ph.CO]}
  \BibitemShut {NoStop}%
\bibitem [{\citenamefont {De~Luca}\ \emph {et~al.}(2020)\citenamefont
  {De~Luca}, \citenamefont {Franciolini}, \citenamefont {Pani},\ and\
  \citenamefont {Riotto}}]{DeLuca:2020qqa}%
  \BibitemOpen
  \bibfield  {author} {\bibinfo {author} {\bibfnamefont {V.}~\bibnamefont
  {De~Luca}}, \bibinfo {author} {\bibfnamefont {G.}~\bibnamefont
  {Franciolini}}, \bibinfo {author} {\bibfnamefont {P.}~\bibnamefont {Pani}}, \
  and\ \bibinfo {author} {\bibfnamefont {A.}~\bibnamefont {Riotto}},\ }\href
  {\doibase 10.1088/1475-7516/2020/06/044} {\bibfield  {journal} {\bibinfo
  {journal} {J. Cosmol. Astropart. Phys.}\ }\textbf {\bibinfo {volume} {06}},\
  \bibinfo {pages} {044} (\bibinfo {year} {2020})},\ \Eprint
  {http://arxiv.org/abs/2005.05641} {arXiv:2005.05641 [astro-ph.CO]}
  \BibitemShut {NoStop}%
\bibitem [{\citenamefont {Franciolini}\ \emph {et~al.}(2022)\citenamefont
  {Franciolini}, \citenamefont {Cotesta}, \citenamefont {Loutrel},
  \citenamefont {Berti}, \citenamefont {Pani},\ and\ \citenamefont
  {Riotto}}]{Franciolini:2021xbq}%
  \BibitemOpen
  \bibfield  {author} {\bibinfo {author} {\bibfnamefont {G.}~\bibnamefont
  {Franciolini}}, \bibinfo {author} {\bibfnamefont {R.}~\bibnamefont
  {Cotesta}}, \bibinfo {author} {\bibfnamefont {N.}~\bibnamefont {Loutrel}},
  \bibinfo {author} {\bibfnamefont {E.}~\bibnamefont {Berti}}, \bibinfo
  {author} {\bibfnamefont {P.}~\bibnamefont {Pani}}, \ and\ \bibinfo {author}
  {\bibfnamefont {A.}~\bibnamefont {Riotto}},\ }\href {\doibase
  10.1103/PhysRevD.105.063510} {\bibfield  {journal} {\bibinfo  {journal}
  {Phys. Rev. D}\ }\textbf {\bibinfo {volume} {105}},\ \bibinfo {pages}
  {063510} (\bibinfo {year} {2022})},\ \Eprint
  {http://arxiv.org/abs/2112.10660} {arXiv:2112.10660 [astro-ph.CO]}
  \BibitemShut {NoStop}%
\bibitem [{\citenamefont {Volonteri}\ \emph {et~al.}(2021)\citenamefont
  {Volonteri}, \citenamefont {Habouzit},\ and\ \citenamefont
  {Colpi}}]{Volonteri:2021sfo}%
  \BibitemOpen
  \bibfield  {author} {\bibinfo {author} {\bibfnamefont {M.}~\bibnamefont
  {Volonteri}}, \bibinfo {author} {\bibfnamefont {M.}~\bibnamefont {Habouzit}},
  \ and\ \bibinfo {author} {\bibfnamefont {M.}~\bibnamefont {Colpi}},\ }\href
  {\doibase 10.1038/s42254-021-00364-9} {\bibfield  {journal} {\bibinfo
  {journal} {Nature Rev. Phys.}\ }\textbf {\bibinfo {volume} {3}},\ \bibinfo
  {pages} {732} (\bibinfo {year} {2021})},\ \Eprint
  {http://arxiv.org/abs/2110.10175} {arXiv:2110.10175 [astro-ph.GA]}
  \BibitemShut {NoStop}%
\bibitem [{\citenamefont {Vitale}\ and\ \citenamefont
  {Evans}(2017)}]{Vitale:2016icu}%
  \BibitemOpen
  \bibfield  {author} {\bibinfo {author} {\bibfnamefont {S.}~\bibnamefont
  {Vitale}}\ and\ \bibinfo {author} {\bibfnamefont {M.}~\bibnamefont {Evans}},\
  }\href {\doibase 10.1103/PhysRevD.95.064052} {\bibfield  {journal} {\bibinfo
  {journal} {Phys. Rev. D}\ }\textbf {\bibinfo {volume} {95}},\ \bibinfo
  {pages} {064052} (\bibinfo {year} {2017})},\ \Eprint
  {http://arxiv.org/abs/1610.06917} {arXiv:1610.06917 [gr-qc]} \BibitemShut
  {NoStop}%
\bibitem [{\citenamefont {Vitale}\ and\ \citenamefont
  {Whittle}(2018)}]{Vitale:2018nif}%
  \BibitemOpen
  \bibfield  {author} {\bibinfo {author} {\bibfnamefont {S.}~\bibnamefont
  {Vitale}}\ and\ \bibinfo {author} {\bibfnamefont {C.}~\bibnamefont
  {Whittle}},\ }\href {\doibase 10.1103/PhysRevD.98.024029} {\bibfield
  {journal} {\bibinfo  {journal} {Phys. Rev. D}\ }\textbf {\bibinfo {volume}
  {98}},\ \bibinfo {pages} {024029} (\bibinfo {year} {2018})},\ \Eprint
  {http://arxiv.org/abs/1804.07866} {arXiv:1804.07866 [gr-qc]} \BibitemShut
  {NoStop}%
\bibitem [{\citenamefont {Ng}\ \emph {et~al.}(2022{\natexlab{a}})\citenamefont
  {Ng}, \citenamefont {Chen}, \citenamefont {Goncharov}, \citenamefont
  {Dupletsa}, \citenamefont {Borhanian}, \citenamefont {Branchesi},
  \citenamefont {Harms}, \citenamefont {Maggiore}, \citenamefont
  {Sathyaprakash},\ and\ \citenamefont {Vitale}}]{Ng:2021sqn}%
  \BibitemOpen
  \bibfield  {author} {\bibinfo {author} {\bibfnamefont {K.~K.~Y.}\
  \bibnamefont {Ng}}, \bibinfo {author} {\bibfnamefont {S.}~\bibnamefont
  {Chen}}, \bibinfo {author} {\bibfnamefont {B.}~\bibnamefont {Goncharov}},
  \bibinfo {author} {\bibfnamefont {U.}~\bibnamefont {Dupletsa}}, \bibinfo
  {author} {\bibfnamefont {S.}~\bibnamefont {Borhanian}}, \bibinfo {author}
  {\bibfnamefont {M.}~\bibnamefont {Branchesi}}, \bibinfo {author}
  {\bibfnamefont {J.}~\bibnamefont {Harms}}, \bibinfo {author} {\bibfnamefont
  {M.}~\bibnamefont {Maggiore}}, \bibinfo {author} {\bibfnamefont {B.~S.}\
  \bibnamefont {Sathyaprakash}}, \ and\ \bibinfo {author} {\bibfnamefont
  {S.}~\bibnamefont {Vitale}},\ }\href {\doibase 10.3847/2041-8213/ac6bea}
  {\bibfield  {journal} {\bibinfo  {journal} {Astrophys. J. Lett.}\ }\textbf
  {\bibinfo {volume} {931}},\ \bibinfo {pages} {L12} (\bibinfo {year}
  {2022}{\natexlab{a}})},\ \Eprint {http://arxiv.org/abs/2108.07276}
  {arXiv:2108.07276 [astro-ph.CO]} \BibitemShut {NoStop}%
\bibitem [{\citenamefont {Ng}\ \emph {et~al.}(2023)\citenamefont {Ng} \emph
  {et~al.}}]{Ng:2022vbz}%
  \BibitemOpen
  \bibfield  {author} {\bibinfo {author} {\bibfnamefont {K.~K.~Y.}\
  \bibnamefont {Ng}} \emph {et~al.},\ }\href {\doibase
  10.1103/PhysRevD.107.024041} {\bibfield  {journal} {\bibinfo  {journal}
  {Phys. Rev. D}\ }\textbf {\bibinfo {volume} {107}},\ \bibinfo {pages}
  {024041} (\bibinfo {year} {2023})},\ \Eprint
  {http://arxiv.org/abs/2210.03132} {arXiv:2210.03132 [astro-ph.CO]}
  \BibitemShut {NoStop}%
\bibitem [{\citenamefont {Hall}\ and\ \citenamefont
  {Evans}(2019)}]{Hall:2019xmm}%
  \BibitemOpen
  \bibfield  {author} {\bibinfo {author} {\bibfnamefont {E.~D.}\ \bibnamefont
  {Hall}}\ and\ \bibinfo {author} {\bibfnamefont {M.}~\bibnamefont {Evans}},\
  }\href {\doibase 10.1088/1361-6382/ab41d6} {\bibfield  {journal} {\bibinfo
  {journal} {Class. Quant. Grav.}\ }\textbf {\bibinfo {volume} {36}},\ \bibinfo
  {pages} {225002} (\bibinfo {year} {2019})},\ \Eprint
  {http://arxiv.org/abs/1902.09485} {arXiv:1902.09485 [astro-ph.IM]}
  \BibitemShut {NoStop}%
\bibitem [{\citenamefont {Iacovelli}\ \emph
  {et~al.}(2022{\natexlab{b}})\citenamefont {Iacovelli}, \citenamefont
  {Mancarella}, \citenamefont {Foffa},\ and\ \citenamefont
  {Maggiore}}]{Iacovelli:2022mbg}%
  \BibitemOpen
  \bibfield  {author} {\bibinfo {author} {\bibfnamefont {F.}~\bibnamefont
  {Iacovelli}}, \bibinfo {author} {\bibfnamefont {M.}~\bibnamefont
  {Mancarella}}, \bibinfo {author} {\bibfnamefont {S.}~\bibnamefont {Foffa}}, \
  and\ \bibinfo {author} {\bibfnamefont {M.}~\bibnamefont {Maggiore}},\ }\href
  {\doibase 10.3847/1538-4365/ac9129} {\bibfield  {journal} {\bibinfo
  {journal} {Astrophys. J. Supp.}\ }\textbf {\bibinfo {volume} {263}},\
  \bibinfo {pages} {2} (\bibinfo {year} {2022}{\natexlab{b}})},\ \Eprint
  {http://arxiv.org/abs/2207.06910} {arXiv:2207.06910 [astro-ph.IM]}
  \BibitemShut {NoStop}%
\bibitem [{\citenamefont {Cutler}\ and\ \citenamefont
  {Flanagan}(1994)}]{Cutler:1994ys}%
  \BibitemOpen
  \bibfield  {author} {\bibinfo {author} {\bibfnamefont {C.}~\bibnamefont
  {Cutler}}\ and\ \bibinfo {author} {\bibfnamefont {E.~E.}\ \bibnamefont
  {Flanagan}},\ }\href {\doibase 10.1103/PhysRevD.49.2658} {\bibfield
  {journal} {\bibinfo  {journal} {Phys. Rev. D}\ }\textbf {\bibinfo {volume}
  {49}},\ \bibinfo {pages} {2658} (\bibinfo {year} {1994})},\ \Eprint
  {http://arxiv.org/abs/gr-qc/9402014} {arXiv:gr-qc/9402014} \BibitemShut
  {NoStop}%
\bibitem [{\citenamefont {Vallisneri}(2008)}]{Vallisneri:2007ev}%
  \BibitemOpen
  \bibfield  {author} {\bibinfo {author} {\bibfnamefont {M.}~\bibnamefont
  {Vallisneri}},\ }\href {\doibase 10.1103/PhysRevD.77.042001} {\bibfield
  {journal} {\bibinfo  {journal} {Phys. Rev. D}\ }\textbf {\bibinfo {volume}
  {77}},\ \bibinfo {pages} {042001} (\bibinfo {year} {2008})},\ \Eprint
  {http://arxiv.org/abs/gr-qc/0703086} {arXiv:gr-qc/0703086} \BibitemShut
  {NoStop}%
\bibitem [{\citenamefont {Rodriguez}\ \emph {et~al.}(2013)\citenamefont
  {Rodriguez}, \citenamefont {Farr}, \citenamefont {Farr},\ and\ \citenamefont
  {Mandel}}]{Rodriguez:2013mla}%
  \BibitemOpen
  \bibfield  {author} {\bibinfo {author} {\bibfnamefont {C.~L.}\ \bibnamefont
  {Rodriguez}}, \bibinfo {author} {\bibfnamefont {B.}~\bibnamefont {Farr}},
  \bibinfo {author} {\bibfnamefont {W.~M.}\ \bibnamefont {Farr}}, \ and\
  \bibinfo {author} {\bibfnamefont {I.}~\bibnamefont {Mandel}},\ }\href
  {\doibase 10.1103/PhysRevD.88.084013} {\bibfield  {journal} {\bibinfo
  {journal} {Phys. Rev. D}\ }\textbf {\bibinfo {volume} {88}},\ \bibinfo
  {pages} {084013} (\bibinfo {year} {2013})},\ \Eprint
  {http://arxiv.org/abs/1308.1397} {arXiv:1308.1397 [astro-ph.IM]} \BibitemShut
  {NoStop}%
\bibitem [{\citenamefont {Dupletsa}\ \emph {et~al.}(2023)\citenamefont
  {Dupletsa}, \citenamefont {Harms}, \citenamefont {Banerjee}, \citenamefont
  {Branchesi}, \citenamefont {Goncharov}, \citenamefont {Maselli},
  \citenamefont {Oliveira}, \citenamefont {Ronchini},\ and\ \citenamefont
  {Tissino}}]{Dupletsa:2022scg}%
  \BibitemOpen
  \bibfield  {author} {\bibinfo {author} {\bibfnamefont {U.}~\bibnamefont
  {Dupletsa}}, \bibinfo {author} {\bibfnamefont {J.}~\bibnamefont {Harms}},
  \bibinfo {author} {\bibfnamefont {B.}~\bibnamefont {Banerjee}}, \bibinfo
  {author} {\bibfnamefont {M.}~\bibnamefont {Branchesi}}, \bibinfo {author}
  {\bibfnamefont {B.}~\bibnamefont {Goncharov}}, \bibinfo {author}
  {\bibfnamefont {A.}~\bibnamefont {Maselli}}, \bibinfo {author} {\bibfnamefont
  {A.~C.~S.}\ \bibnamefont {Oliveira}}, \bibinfo {author} {\bibfnamefont
  {S.}~\bibnamefont {Ronchini}}, \ and\ \bibinfo {author} {\bibfnamefont
  {J.}~\bibnamefont {Tissino}},\ }\href {\doibase 10.1016/j.ascom.2022.100671}
  {\bibfield  {journal} {\bibinfo  {journal} {Astron. Comput.}\ }\textbf
  {\bibinfo {volume} {42}},\ \bibinfo {pages} {100671} (\bibinfo {year}
  {2023})},\ \Eprint {http://arxiv.org/abs/2205.02499} {arXiv:2205.02499
  [gr-qc]} \BibitemShut {NoStop}%
\bibitem [{\citenamefont {Borhanian}(2021)}]{Borhanian:2020ypi}%
  \BibitemOpen
  \bibfield  {author} {\bibinfo {author} {\bibfnamefont {S.}~\bibnamefont
  {Borhanian}},\ }\href {\doibase 10.1088/1361-6382/ac1618} {\bibfield
  {journal} {\bibinfo  {journal} {Class. Quant. Grav.}\ }\textbf {\bibinfo
  {volume} {38}},\ \bibinfo {pages} {175014} (\bibinfo {year} {2021})},\
  \Eprint {http://arxiv.org/abs/2010.15202} {arXiv:2010.15202 [gr-qc]}
  \BibitemShut {NoStop}%
\bibitem [{\citenamefont {London}\ \emph {et~al.}(2018)\citenamefont {London},
  \citenamefont {Khan}, \citenamefont {Fauchon-Jones}, \citenamefont
  {Garc\'\i{}a}, \citenamefont {Hannam}, \citenamefont {Husa}, \citenamefont
  {Jim\'enez-Forteza}, \citenamefont {Kalaghatgi}, \citenamefont {Ohme},\ and\
  \citenamefont {Pannarale}}]{London:2017bcn}%
  \BibitemOpen
  \bibfield  {author} {\bibinfo {author} {\bibfnamefont {L.}~\bibnamefont
  {London}}, \bibinfo {author} {\bibfnamefont {S.}~\bibnamefont {Khan}},
  \bibinfo {author} {\bibfnamefont {E.}~\bibnamefont {Fauchon-Jones}}, \bibinfo
  {author} {\bibfnamefont {C.}~\bibnamefont {Garc\'\i{}a}}, \bibinfo {author}
  {\bibfnamefont {M.}~\bibnamefont {Hannam}}, \bibinfo {author} {\bibfnamefont
  {S.}~\bibnamefont {Husa}}, \bibinfo {author} {\bibfnamefont {X.}~\bibnamefont
  {Jim\'enez-Forteza}}, \bibinfo {author} {\bibfnamefont {C.}~\bibnamefont
  {Kalaghatgi}}, \bibinfo {author} {\bibfnamefont {F.}~\bibnamefont {Ohme}}, \
  and\ \bibinfo {author} {\bibfnamefont {F.}~\bibnamefont {Pannarale}},\ }\href
  {\doibase 10.1103/PhysRevLett.120.161102} {\bibfield  {journal} {\bibinfo
  {journal} {Phys. Rev. Lett.}\ }\textbf {\bibinfo {volume} {120}},\ \bibinfo
  {pages} {161102} (\bibinfo {year} {2018})},\ \Eprint
  {http://arxiv.org/abs/1708.00404} {arXiv:1708.00404 [gr-qc]} \BibitemShut
  {NoStop}%
\bibitem [{\citenamefont {Aghanim}\ \emph {et~al.}(2020)\citenamefont {Aghanim}
  \emph {et~al.}}]{Planck:2018vyg}%
  \BibitemOpen
  \bibfield  {author} {\bibinfo {author} {\bibfnamefont {N.}~\bibnamefont
  {Aghanim}} \emph {et~al.} (\bibinfo {collaboration} {Planck}),\ }\href
  {\doibase 10.1051/0004-6361/201833910} {\bibfield  {journal} {\bibinfo
  {journal} {Astron. Astrophys.}\ }\textbf {\bibinfo {volume} {641}},\ \bibinfo
  {pages} {A6} (\bibinfo {year} {2020})},\ \bibinfo {note} {[Erratum:
  Astron.Astrophys. 652, C4 (2021)]},\ \Eprint
  {http://arxiv.org/abs/1807.06209} {arXiv:1807.06209 [astro-ph.CO]}
  \BibitemShut {NoStop}%
\bibitem [{\citenamefont {Hogg}(1999)}]{Hogg:1999ad}%
  \BibitemOpen
  \bibfield  {author} {\bibinfo {author} {\bibfnamefont {D.~W.}\ \bibnamefont
  {Hogg}},\ }\href@noop {} {\  (\bibinfo {year} {1999})},\ \Eprint
  {http://arxiv.org/abs/astro-ph/9905116} {arXiv:astro-ph/9905116} \BibitemShut
  {NoStop}%
\bibitem [{\citenamefont {Dominik}\ \emph {et~al.}(2015)\citenamefont
  {Dominik}, \citenamefont {Berti}, \citenamefont {O'Shaughnessy},
  \citenamefont {Mandel}, \citenamefont {Belczynski}, \citenamefont {Fryer},
  \citenamefont {Holz}, \citenamefont {Bulik},\ and\ \citenamefont
  {Pannarale}}]{Dominik:2014yma}%
  \BibitemOpen
  \bibfield  {author} {\bibinfo {author} {\bibfnamefont {M.}~\bibnamefont
  {Dominik}}, \bibinfo {author} {\bibfnamefont {E.}~\bibnamefont {Berti}},
  \bibinfo {author} {\bibfnamefont {R.}~\bibnamefont {O'Shaughnessy}}, \bibinfo
  {author} {\bibfnamefont {I.}~\bibnamefont {Mandel}}, \bibinfo {author}
  {\bibfnamefont {K.}~\bibnamefont {Belczynski}}, \bibinfo {author}
  {\bibfnamefont {C.}~\bibnamefont {Fryer}}, \bibinfo {author} {\bibfnamefont
  {D.~E.}\ \bibnamefont {Holz}}, \bibinfo {author} {\bibfnamefont
  {T.}~\bibnamefont {Bulik}}, \ and\ \bibinfo {author} {\bibfnamefont
  {F.}~\bibnamefont {Pannarale}},\ }\href {\doibase
  10.1088/0004-637X/806/2/263} {\bibfield  {journal} {\bibinfo  {journal}
  {Astrophys. J.}\ }\textbf {\bibinfo {volume} {806}},\ \bibinfo {pages} {263}
  (\bibinfo {year} {2015})},\ \Eprint {http://arxiv.org/abs/1405.7016}
  {arXiv:1405.7016 [astro-ph.HE]} \BibitemShut {NoStop}%
\bibitem [{\citenamefont {Wilk}\ and\ \citenamefont
  {Gnanadesikan}(1968)}]{gnanadesikan1968probability}%
  \BibitemOpen
  \bibfield  {author} {\bibinfo {author} {\bibfnamefont {M.~B.}\ \bibnamefont
  {Wilk}}\ and\ \bibinfo {author} {\bibfnamefont {R.}~\bibnamefont
  {Gnanadesikan}},\ }\href@noop {} {\bibfield  {journal} {\bibinfo  {journal}
  {Biometrika}\ }\textbf {\bibinfo {volume} {55}},\ \bibinfo {pages} {1}
  (\bibinfo {year} {1968})}\BibitemShut {NoStop}%
\bibitem [{\citenamefont {Abbott}\ \emph {et~al.}(2016)\citenamefont {Abbott}
  \emph {et~al.}}]{LIGOScientific:2016aoc}%
  \BibitemOpen
  \bibfield  {author} {\bibinfo {author} {\bibfnamefont {B.~P.}\ \bibnamefont
  {Abbott}} \emph {et~al.} (\bibinfo {collaboration} {LIGO Scientific,
  Virgo}),\ }\href {\doibase 10.1103/PhysRevLett.116.061102} {\bibfield
  {journal} {\bibinfo  {journal} {Phys. Rev. Lett.}\ }\textbf {\bibinfo
  {volume} {116}},\ \bibinfo {pages} {061102} (\bibinfo {year} {2016})},\
  \Eprint {http://arxiv.org/abs/1602.03837} {arXiv:1602.03837 [gr-qc]}
  \BibitemShut {NoStop}%
\bibitem [{\citenamefont {Abbott}\ \emph
  {et~al.}(2021{\natexlab{c}})\citenamefont {Abbott} \emph
  {et~al.}}]{LIGOScientific:2021psn}%
  \BibitemOpen
  \bibfield  {author} {\bibinfo {author} {\bibfnamefont {R.}~\bibnamefont
  {Abbott}} \emph {et~al.} (\bibinfo {collaboration} {LIGO Scientific, Virgo,
  KAGRA}),\ }\href@noop {} {\  (\bibinfo {year} {2021}{\natexlab{c}})},\
  \Eprint {http://arxiv.org/abs/2111.03634} {arXiv:2111.03634 [astro-ph.HE]}
  \BibitemShut {NoStop}%
\bibitem [{\citenamefont {Schutz}(2011)}]{Schutz:2011tw}%
  \BibitemOpen
  \bibfield  {author} {\bibinfo {author} {\bibfnamefont {B.~F.}\ \bibnamefont
  {Schutz}},\ }\href {\doibase 10.1088/0264-9381/28/12/125023} {\bibfield
  {journal} {\bibinfo  {journal} {Class. Quant. Grav.}\ }\textbf {\bibinfo
  {volume} {28}},\ \bibinfo {pages} {125023} (\bibinfo {year} {2011})},\
  \Eprint {http://arxiv.org/abs/1102.5421} {arXiv:1102.5421 [astro-ph.IM]}
  \BibitemShut {NoStop}%
\bibitem [{\citenamefont {Koushiappas}\ and\ \citenamefont
  {Loeb}(2017)}]{Koushiappas:2017kqm}%
  \BibitemOpen
  \bibfield  {author} {\bibinfo {author} {\bibfnamefont {S.~M.}\ \bibnamefont
  {Koushiappas}}\ and\ \bibinfo {author} {\bibfnamefont {A.}~\bibnamefont
  {Loeb}},\ }\href {\doibase 10.1103/PhysRevLett.119.221104} {\bibfield
  {journal} {\bibinfo  {journal} {Phys. Rev. Lett.}\ }\textbf {\bibinfo
  {volume} {119}},\ \bibinfo {pages} {221104} (\bibinfo {year} {2017})},\
  \Eprint {http://arxiv.org/abs/1708.07380} {arXiv:1708.07380 [astro-ph.CO]}
  \BibitemShut {NoStop}%
\bibitem [{\citenamefont {Franciolini}\ \emph {et~al.}(2023)\citenamefont
  {Franciolini}, \citenamefont {Iacovelli}, \citenamefont {Mancarella},
  \citenamefont {Maggiore}, \citenamefont {Pani},\ and\ \citenamefont
  {Riotto}}]{Franciolini:2023opt}%
  \BibitemOpen
  \bibfield  {author} {\bibinfo {author} {\bibfnamefont {G.}~\bibnamefont
  {Franciolini}}, \bibinfo {author} {\bibfnamefont {F.}~\bibnamefont
  {Iacovelli}}, \bibinfo {author} {\bibfnamefont {M.}~\bibnamefont
  {Mancarella}}, \bibinfo {author} {\bibfnamefont {M.}~\bibnamefont
  {Maggiore}}, \bibinfo {author} {\bibfnamefont {P.}~\bibnamefont {Pani}}, \
  and\ \bibinfo {author} {\bibfnamefont {A.}~\bibnamefont {Riotto}},\
  }\href@noop {} {\  (\bibinfo {year} {2023})},\ \Eprint
  {http://arxiv.org/abs/2304.03160} {arXiv:2304.03160 [gr-qc]} \BibitemShut
  {NoStop}%
\bibitem [{\citenamefont {Moore}\ and\ \citenamefont
  {Gerosa}(2021)}]{Moore:2021xhn}%
  \BibitemOpen
  \bibfield  {author} {\bibinfo {author} {\bibfnamefont {C.~J.}\ \bibnamefont
  {Moore}}\ and\ \bibinfo {author} {\bibfnamefont {D.}~\bibnamefont {Gerosa}},\
  }\href {\doibase 10.1103/PhysRevD.104.083008} {\bibfield  {journal} {\bibinfo
   {journal} {Phys. Rev. D}\ }\textbf {\bibinfo {volume} {104}},\ \bibinfo
  {pages} {083008} (\bibinfo {year} {2021})},\ \Eprint
  {http://arxiv.org/abs/2108.02462} {arXiv:2108.02462 [gr-qc]} \BibitemShut
  {NoStop}%
\bibitem [{\citenamefont {Ng}\ \emph {et~al.}(2022{\natexlab{b}})\citenamefont
  {Ng}, \citenamefont {Franciolini}, \citenamefont {Berti}, \citenamefont
  {Pani}, \citenamefont {Riotto},\ and\ \citenamefont {Vitale}}]{Ng:2022agi}%
  \BibitemOpen
  \bibfield  {author} {\bibinfo {author} {\bibfnamefont {K.~K.~Y.}\
  \bibnamefont {Ng}}, \bibinfo {author} {\bibfnamefont {G.}~\bibnamefont
  {Franciolini}}, \bibinfo {author} {\bibfnamefont {E.}~\bibnamefont {Berti}},
  \bibinfo {author} {\bibfnamefont {P.}~\bibnamefont {Pani}}, \bibinfo {author}
  {\bibfnamefont {A.}~\bibnamefont {Riotto}}, \ and\ \bibinfo {author}
  {\bibfnamefont {S.}~\bibnamefont {Vitale}},\ }\href {\doibase
  10.3847/2041-8213/ac7aae} {\bibfield  {journal} {\bibinfo  {journal}
  {Astrophys. J. Lett.}\ }\textbf {\bibinfo {volume} {933}},\ \bibinfo {pages}
  {L41} (\bibinfo {year} {2022}{\natexlab{b}})},\ \Eprint
  {http://arxiv.org/abs/2204.11864} {arXiv:2204.11864 [astro-ph.CO]}
  \BibitemShut {NoStop}%
\bibitem [{\citenamefont {Mukherjee}\ \emph {et~al.}(2022)\citenamefont
  {Mukherjee}, \citenamefont {Meinema},\ and\ \citenamefont
  {Silk}}]{Mukherjee:2021itf}%
  \BibitemOpen
  \bibfield  {author} {\bibinfo {author} {\bibfnamefont {S.}~\bibnamefont
  {Mukherjee}}, \bibinfo {author} {\bibfnamefont {M.~S.~P.}\ \bibnamefont
  {Meinema}}, \ and\ \bibinfo {author} {\bibfnamefont {J.}~\bibnamefont
  {Silk}},\ }\href {\doibase 10.1093/mnras/stab3756} {\bibfield  {journal}
  {\bibinfo  {journal} {Mon. Not. Roy. Astron. Soc.}\ }\textbf {\bibinfo
  {volume} {510}},\ \bibinfo {pages} {6218} (\bibinfo {year} {2022})},\ \Eprint
  {http://arxiv.org/abs/2107.02181} {arXiv:2107.02181 [astro-ph.CO]}
  \BibitemShut {NoStop}%
\bibitem [{\citenamefont {Bavera}\ \emph {et~al.}(2022)\citenamefont {Bavera},
  \citenamefont {Franciolini}, \citenamefont {Cusin}, \citenamefont {Riotto},
  \citenamefont {Zevin},\ and\ \citenamefont {Fragos}}]{Bavera:2021wmw}%
  \BibitemOpen
  \bibfield  {author} {\bibinfo {author} {\bibfnamefont {S.~S.}\ \bibnamefont
  {Bavera}}, \bibinfo {author} {\bibfnamefont {G.}~\bibnamefont {Franciolini}},
  \bibinfo {author} {\bibfnamefont {G.}~\bibnamefont {Cusin}}, \bibinfo
  {author} {\bibfnamefont {A.}~\bibnamefont {Riotto}}, \bibinfo {author}
  {\bibfnamefont {M.}~\bibnamefont {Zevin}}, \ and\ \bibinfo {author}
  {\bibfnamefont {T.}~\bibnamefont {Fragos}},\ }\href {\doibase
  10.1051/0004-6361/202142208} {\bibfield  {journal} {\bibinfo  {journal}
  {Astron. Astrophys.}\ }\textbf {\bibinfo {volume} {660}},\ \bibinfo {pages}
  {A26} (\bibinfo {year} {2022})},\ \Eprint {http://arxiv.org/abs/2109.05836}
  {arXiv:2109.05836 [astro-ph.CO]} \BibitemShut {NoStop}%
\end{thebibliography}%

\end{document}